
\magnification=1200
\hsize=13cm
\def\newline{\hfil\break}  
\def\proclaim#1#2{\medskip\noindent{\bf #1}\quad \begingroup #2}
\def\endproclaim{\endgroup\medskip}

\topskip10pt plus40pt
\def\title#1{\medskip\penalty-30\noindent{\bf #1}\quad}
\def\endtitle{\par \penalty30\medskip \penalty30}
\clubpenalty=30

\headline={\hfil}
\footline={\hss\tenrm\folio\hss}

{\bf
\centerline{Neural Unpredictability, The Interpretation Of
Quantum Theory,}
\centerline{And The Mind-Body Problem. \ 
\footnote{}{\tenrm * quant-ph/0208033, August 2002.}}
\medskip

\centerline{Matthew J. Donald}
\medskip

\centerline{The Cavendish Laboratory,  Madingley Road,}  

\centerline{Cambridge CB3 0HE,  Great Britain.}
\smallskip

\centerline{ e-mail: \quad matthew.donald@phy.cam.ac.uk}
\smallskip
{\hfill web site: \quad 
{\catcode`\~=12 \catcode`\q=9 http://www.poco.phy.cam.ac.uk/q~mjd1014}}
\hfill}
\medskip

\proclaim{abstract}{}  It has been suggested, on the one hand, that
quantum states are just states of knowledge; and, on the other, that
quantum theory is merely a theory of correlations.  These suggestions
are confronted with problems about the nature of psycho-physical
parallelism and about how we could define probabilities for our
individual future observations given our individual present and
previous observations.  The complexity of the problems is underlined
by arguments that unpredictability in ordinary everyday neural
functioning, ultimately stemming from small-scale uncertainties in
molecular motions, may overwhelm, by many orders of magnitude,
many conventionally recognized sources of observed ``quantum''
uncertainty.  Some possible ways of avoiding the problems are
considered but found wanting.  It is proposed that a complete
understanding of the relationship between subjective experience and
its physical correlates requires the introduction of mathematical
definitions and indeed of new physical laws.
\endproclaim

\title{Plausible Ideas Confronted.}
\endtitle

Recently, some plausible ideas about quantum theory have led to
claims about the interpretation of the theory which, in my
opinion, are simplistic.  On the one hand, it has, from time to time,
been suggested that quantum states are merely states of knowledge
(or of belief) (Wolfe 1936, Wigner 1961, Peierls 1991, Fuchs 2002). 
This idea has led to the claim that quantum theory ``needs no
interpretation'' (Fuchs and Peres 2000).  On the other hand, various
authors have argued, in various ways, that quantum theory is
fundamentally just a theory of relations or of correlations (Wheeler
1957, Saunders 1995, 1998, Rovelli 1996, Mermin 1998).  This idea
has led to the claim that it is not necessary in a many-worlds
interpretation to specify the concept of a ``world'' (Wallace 2001a,
2001b; an introductory account is given by Butterfield 2001).  In this
paper, I shall confront these ideas with some empirical facts about the
complexity and unpredictability of human neural processing which
indicate how wrong it would be to think of human observers as
deterministic robots.  I shall argue that these facts should prevent us
from being satisfied with an imprecise approach to the problems of
understanding the composition and possible changes, either of our
knowledge or of our correlations.

\title{The Wider Picture.}
\endtitle

The wider purpose of this paper is to draw attention to problems
which could be significant in any attempt to understand the physical
underpinning of individual human awarenesses.  In a long series of
previous papers (Donald 1990, 1992, 1995, 1997, 1999), I have
proposed a solution to these problems in the context of a many-minds
interpretation of quantum theory.  Whatever else they might amount
to, I believe that these papers do provide a serious attempt to grapple
with the kind of detailed technical issues which are involved in
a complete and consistent version of this sort of interpretation.  In
this paper therefore, I shall use my earlier work to illustrate some of
the ways in which difficulties can arise and some of the ways in
which those difficulties might be resolved.  In particular, much of
this paper relates to the problem of giving a description of the
physical structure of an individual human observer which is
sufficiently complete to express everything about that structure
which is directly relevant to the observer's mental life, and so I shall
review my previous attempts to find the most simple such
description.  I shall also frequently refer to ideas from consistent
histories theory which I see as a useful step towards a more complete
analysis.  Nevertheless, many of the broader arguments here do not
depend on the assumption of a universal (Everettian) quantum
theory; albeit that some of the problems may be least easy to ignore
given the radical indeterminism of that kind of theory.

The fundamental premise is that it is not an illusion to suppose that an
individual consciousness has a past and a future, or a range of possible
futures.  Given that premise, we can ask what can be said about
someone's possible future experiences and their probabilities.  It will
be taken for granted here that we have full knowledge of the
observer-independent behaviour of all the physical matter involved,
and it will also, of course, be assumed that mind has no direct
physical effect on matter.  However, in a many-minds interpretation,
the observer-independent behaviour, given by some form of
Schr\"odinger equation acting without ``collapse'' at the level of the
entire universe, is very different from the apparent behaviour as
observed by any individual.  Even in the context of a classical
deterministic physics, we cannot hope for a complete analysis of the
experiences of an observer unless we know what aspects of the
matter constitute the observer.  With the observed nature of human
neural processing and the observed unpredictability of physical events
at a molecular scale, characterizing the physical constitution of
humans as observers raises difficult problems, however we might
wish to interpret quantum theory.  One of the underlying aims of the
paper, therefore, is to justify the proposal that a complete
understanding of the relationship between subjective experience and
its physical correlates, and of the dynamics of that experience, does
require the introduction of mathematical definitions and indeed of
new physical laws.  In part, this will be on the grounds that the facts
about human observers leave so many openings, that nothing less
specific can be satisfactory.

Such mathematical definitions and new physical laws are central in a
complete many-minds interpretation (cf.~Barrett 1999, p.197).  It
seems to me that, aesthetically at least, this is a great advantage of
interpretations of this type.  At the end of the paper, we shall briefly
consider the situation in some other interpretations; supposing that
the technical problems of those interpretations can be solved.  In those
interpretations, how the private experiences of an observer are
constituted, while still problematic, can be separated from the
physical dynamics.  This allows the mind-body problem to continue
to seem as strange and unnatural as it sometimes appears in the
writings of philosophers ignorant of quantum physics.  In a
many-minds interpretation, however, the constitution of the private
experiences of the observer is essential to the dynamics of those
experiences, so that there is a compatibility between the problems of
the philosophy of mind and those of the philosophy of physics.

\title{Multiply-Localized Indeterminacy and Natural Assumptions.}
\endtitle

According to many interpretations of quantum theory, the observed
unpredicta\-bility of human neural processing is a consequence of the
small scale indeterminacy of molecular motions in a warm wet fluid. 
This indeterminacy appears to be ``multiply-localized'', in the sense
that it seems to involve many distinct individual event choices at
many different places.  In trying to understand psycho-physical
parallelism, or the mind-body problem, in a multiply-localized
indeterministic physics, it seems to be very difficult to avoid
intuitively natural assumptions which, unfortunately, are question
begging.  In particular, indeterministic observations are naturally
interpreted in terms of a set of well-defined futures, each with
associated probabilities.  The problem, even ignoring quantum
complications, is that, in a multiply-localized indeterministic physics,
such an interpretation necessarily involves a particular choice of
boundary of observation.  It may also involve a choice of a scale of
observation.  Without such choices, futures and probabilities for
individual observers are undefined.  In everyday life, however, these
choices are implicit in our nature as observers.  We automatically take
the existence of an observer for granted; we always see from some
``point of view''.

It may be tempting to suppose that all that is being referred to here is
a question of level of coarse-graining and so it may seem that
any problem can be resolved just by reference to a natural consistency
between different levels.  Undoubtedly there can be a natural
consistency between different levels of coarse-graining in simple
classical models, but this is of little relevance.   The systems we are
considering are very very far from simple.  They are also not
perfectly classical and, without careful and specific definition,
different levels cannot be assumed to be better than approximately
consistent.  Most importantly, however, they are self-observing.  A
change in level of observation of a system external to an observer
merely requires taking or discarding additional data; a change in
internal level makes an entirely new observer.  Moreover, even the
task of finding a natural choice of levels of coarse-graining for
internal observation within the human brain seems essentially
equivalent to the task of characterizing the physical structure of
humans as observers.  It is all very well for a relational approach to
probability (Saunders 1998) to suggest that probability should be
thought of as a relation between a present and a possible future.  But
such a relation depends on a present and a future being specified. 
The self-observation of the brain is a lively dynamical process,
continually subject to possibilities of growth and decay which are not
easily classified.

\title{Brains are Physical Systems.}
\endtitle

Some authors have argued that problems relating to the nature of
consciousness are not problems for physics and that they should be
``set aside'' (Mermin 1998); other authors are entirely explicit about
the choices to be left to the (external) observer (Griffiths 1998). 
Although these attitudes seems to me to be far superior to simple
denial that any problems exist, I shall also argue here that they are
ultimately untenable.  The fundamental problem in the interpretation
of quantum theory is to understand the states occupied (or apparently
occupied) at any moment by any physical system.  Brains are physical
systems and so quantum theory calls their states and histories into
question, but brains also appear to be the systems through which all
observations are ultimately made.  Understanding the act of
observation therefore requires an understanding of the physical
nature of neural processing.

\title{Psycho-Physical Parallelism.}
\endtitle

In conventional terms, the evidence for mind-brain parallelism is so
overwhelming that many have argued that it amounts to an identity. 
Everything we experience is directed reflected in the functioning and
structure of our nervous system.  You cannot see the back of your
head without a mirror.  You cannot see at all, but perhaps you can still
move your toes, if your eyes are gouged out or your optic nerve is
cut or the back of your head is shot off.  You cannot move your toes,
but perhaps you can still see, if your foot is amputated or your spinal
cord is cut or you get a bullet in the top of your head.  Indeed, it
seems that the parallels between our physical structure and our
experience can be made as precise and detailed as we like.  Modern
philosophy of mind is concerned either with denying (in the context
of classical physics) that there is anything beyond the physical
structure (Dennett 1991) or with trying to understand the parallelism
(Chalmers 1996).  Quantum physics calls all our ideas of ``physical
structure'' into question and seems to give a special role to the
``observer''.  In this context, therefore it seems sensible, at least at
the outset, not to condemn as mere naivity the separation between
the ideas of mind and of brain.  

\title{Self-Reference.}
\endtitle

Psycho-physical parallelism implies that the knowledge of any
individual must be reflected in his or her physical neural structure or
functioning.  Individual knowledge is the foundation of all knowledge,
and so the physical structure of individual human brains underlies all
knowledge.  As physical systems, brains are merely complex
collections of ions, atoms, and molecules.  Ions, atoms, and molecules
are quantum mechanical systems, and so the brain is a quantum
mechanical system.  If quantum states are states of knowledge then
the physical structure of a brain, considered as a quantum system, is
fundamental to that knowledge.  This might appear to lead to
problems of self-reference.  Indeed, in a published reply to published
correspondence, Fuchs and Peres (2000) write, ``The main point of
disagreement we have with {Brun} and {Griffiths} is about the
existence of a wavefunction of the universe that would include all its
degrees of freedom, even those in our brains. We assert that this
would lead to absurd self-referential paradoxes.''  Of course, if there
really are ``absurd self-referential paradoxes'' here, then they
amount to a reductio ad absurdum for the thesis that quantum states
are states of knowledge, but I am not at all sure that that thesis is
entirely wrong; certainly, it has led to some valuable insights (Fuchs
2001a, Caves, Fuchs, and Schack 2001).  My opinion is that the issue
is problematic as much as paradoxical.  It is not, for example, that an
observer must know everything that could be known about himself
and also have the additional knowledge that he knows it.  Rather, an
observer must, in some very broad sense, {\sl be something\/}
through which his knowledge is experienced.  As far as explicit or
usable knowledge is concerned, that something is merely some kind of
extreme upper bound.  In my many-minds theory, the
possible futures of an observer, and therefore the external quantum
states he can expect to observe, are ultimately determined by what
he is, rather than by his explicit knowledge (see the discussion around
definition 6.4 in Donald 1999).  Nevertheless, in many relevant cases,
the determining information can be expressed in terms of
explicit knowledge in such a way as to provide a foundation for the
idea of quantum states as states of knowledge.  For example, when
Alice has observed an up spin for one spin from a singlet state, that
observation, which is sufficient to determine that she will observe the
other spin to be down, is both part of what she has become and part
of what she explicitly knows.

\title{Internal and External Observers.}
\endtitle

If knowledge is fundamental, then we ought ultimately to be able to
identify the aspects of the structure or state of a brain on which that
knowledge depends.  The empirical facts which we ought to be able
explain do not consist simply of the existence of all sorts of
correlations between all sorts of different physical systems.  The most
significant fact of all is that we do each individually experience a
world, or what we call a world.  How can that fact be expressed in the
mathematical framework of quantum theory?  If we cannot identify
the aspects of a brain on which that fact is based, then our
foundations are built on sand.  Wallace (2001b) has argued that this is
not important.  According to him, worlds in a many-worlds theory or
minds in a many-minds theory are like tigers and a tiger is ``any
pattern which behaves as a tiger''.  The problem with this is that
although ``behaviour'' or ``tigers'' or ``correlations''  or ``knowledge''
can easily be spotted by an observer external to the system being
considered, physical observers are constituted by their own physical
systems. The observers of many-worlds quantum theory, for
example, are internal to the universal quantum state and must, in
some sense, find themselves within it.  An implicit appeal to an
external observer is analogous to Bishop Berkeley's solution to the
problems of idealism by appeal to a divine mind. 

\title{The Entry-Price Problem.}
\endtitle

Another reason why, if knowledge is fundamental, we need to be able
to identify aspects of neural structures as a basis for our actual
observations is that otherwise we cannot define probabilities for our
future observations from our present and previous observations.  In
an indeterministic theory, probabilities are what get us from moment
to moment.  An entirely classical model is sufficient to demonstrate
the problem:  Suppose you are offered the opportunity to play a game
in which five ordered, fair, and independent coins marked 0 on one
side and 1 on the other will be tossed and in which your winning will
be the sum displayed as a binary number.  Thus, with equal
probability, you could win any sum between 00000 = 0 and 11111 =
31.  A fair entry price for the game would be 15.5.  If the first two
coins are to be ignored, then a fair entry price would be 3.5.  But  a
game with an undetermined number of coins has no fair entry price;
neither does a game during which some coins may arbitrarily be
removed or others may be added.  In this sort of undefined game,
even if we could make a list of possible outcomes,  we would still not
have a good theory of temporal change.  

With equiprobable events, the individual events in a five coin game
have probability ${1 \over 32}$, while in a three coin game the
individual events have probability ${1 \over 8}$.  With an external
observer, this is all perfectly simple:  ${1 \over 32}$ is the probability
of seeing 10101, while ${1 \over 8}$ is the probability of the last
three coins reading 101.  But what if the physical structure of the
observer is a set of coins?  Such an observer might say that his
current state was 01101 and that, at the next ``snapshot'', he could be
one of thirty-two equiprobable five coin possibilities; or he might say
that his current state was 01101 and that, at the next ``snapshot'', he
could be one of eight equiprobable three coin possibilities; but surely,
there would be something missing if he said that, at the next
``snapshot'', there were a total of 40 possibilities, either five coin or
three coin, but there was no answer to the question of how likely
each would be?  It is, of course, true that three coins is a subset of five
coins, so that classical probability theory (as long as it is appropriate)
provides a perfectly adequate external description which allows for
choices in the scale of observation.  Nevertheless, once again, we are
not seeking a description subject to the choices of some implicit
external observer, but rather a basis for our actual observations. 

The main goal of this paper is to argue that problems analogous to the
entry-price problem do arise in the analysis of psycho-physical
parallelism and that there are no easy ways of dealing with, or
avoiding, those problems.  In particular, the complexities of the
physical expression of neural information are such that there is no
easy way of specifying either an exact number of ``coins'' in play at
any moment or the possible changes with time in such a number. 
Nevertheless, analogs of such specifications are required if the
``internal'' probabilities of observations are to be determined, and,
because the details of neural processing are so unpredictable, those
probabilities do depend significantly on how that requirement is met.

\title{Quantum Problems.}
\endtitle

The entry-price problem for an undetermined coin game is simple by
comparison with the problems of a fully quantum mechanical theory. 
As an example, one can consider the set selection problem in
consistent histories theory which effectively adds the problem of
defining the coins themselves as quantum entities.  There are
continuously many sets of consistent histories and they are, of course,
not all mutually consistent.  The entry-price problem merely
requires us to choose, at each moment, a classical coarse graining (the
number of coins in play).  For the brain, there are an entity-type
problems, problems of continuous variation for given entity-types,
and time-dependent coarse-graining problems.  It is also possible
with quantum theory that there can be further problems specific to
the nature of the subject.  For example, were the brain to function as a
quantum computer, it might be that we would face problems directly
involving interference effects or requirements on the purity of
particular subsystem states.  However, I do not in fact believe that
this sort of problem does arise.  This is because I do not believe that
the brain has somehow evolved any special capacities which would
make it in any way like a quantum computer.  The arguments
developed here therefore recognize the importance of decoherence
theory for the description of warm wet systems like the human brain,
but aim to demonstrate that, even although that theory may, by and
large, move the mathematics from quantum to approximately
quasi-classical probabilities, it still does not solve the fundamental
problem of how awareness is represented by changing subsystems of
a quantum universe. 

\title{The Insufficiency of Decoherence.}
\endtitle

In this context, it may also be noted that, especially for complex
systems, decoherence theory does not by itself solve the preferred
basis problem, but merely provides a framework within which a
quasi-classical solution to the preferred basis problem is not ruled
out.  Consider, for example, the state of a living human brain
considered as an unobserved or non-collapsing quantum system for a
period of a few minutes or longer.  The blood delivers oxygen to the
brain and carries away carbon dioxide, at a rate, calculated from data
on page 290 of Bell, Emslie-Smith, and Paterson (1980), of the order
of $10^{19}$ molecules per second.  In the process, heat is also
exchanged to maintain the vital constancy of temperature.  This
means that, at least on small scales, the quantum states of many
significant degrees of freedom of the unobserved brain will tend
to approximate to thermal equilibrium states.  The local density
matrices  for those degrees of freedom will therefore be close to
density matrices with multiple degeneracies.  Such density matrices
do have approximate decoherent decompositions into quasi-classical
states, but they also have many other approximate decompositions
which are not quasi-classical (some relevant examples are given in
Donald 1998). 

\title{Knowledge.}
\endtitle

Many of the most difficult questions for interpretations of quantum
theory might seem to be answerable in terms of an observer's
knowledge.  The observer knows in what context an observation takes
place, and what is being measured, and perhaps even what basis he
or she prefers. If a coarse-graining is required, then perhaps it can
be chosen in terms of what the observer can know.  When we do try to
assign a quantum state to an external physical system, we do often
make reference to our knowledge and we often face choices which
seem knowledge-related.  For example, with a sample of gas, we
might assign a state close to a Gibbs' equilibrium state given the
``known'' temperature and pressure.  On the other hand, we may
think of the ``real'' state as being some ``unknown'' wavefunction
(i.e.~pure state) in a Hilbert space for the atomic particles of the gas. 
Similarly, there are different ways in which quantum states might be
assigned to a functioning human brain and these ways may also be
knowledge-related.  The knowledge of central interest here is the
knowledge of the brain's possessor.  The mind-brain parallelism
discussed above suggests that this knowledge be specified in
functional or biological terms.

\title{The Basis of Neural Functioning.}
\endtitle

Neural firing appears to be the basis of neural functioning.  There are
around $10^{11}$ neurons in a human brain.  At least as a first
approximation, biologically important information in the brain seems
to be coded into the all-or-nothing dichotomies for individual
neurons of either firing or not-firing.  If we want to assign quantum
states to our own brains compatible with our own knowledge, then it
might seem that a reasonable starting point would be to assume that
the current pattern of neural firing is ``known''. 

\title{What Do We Know?}
\endtitle  

To say that we ``know'' about the state of our own brain is not of
course to say that we have direct scientific knowledge of our own
neurophysiology.  Rather, we know the state through the
representations we construct of external reality.  Our current
experiences are what they are entirely because of our present and
previous brain states.  These states are apparently caused by the
external world and that is {\sl how }we experience them, but {\sl
what }we experience is the brain states.  Yet it is difficult to see
what level of detail our knowledge provides.  When, for example, we
watch fireworks, we know increased levels of excitation in different
parts of our head as bright lights and loud noises.  However it seems
implausible that our awareness of an external world would be
sufficient to determine well-localized positions for each individual
molecule inside the brain.  On the other hand, while it might seem
desirable to try to define a correct level of coarse-graining in terms of
a level of phenomenological knowledge much coarser than that of
detailed neural firing patterns, it is hard to find any straightforward
way in which this can be done.  Furthermore, any problems of
specification and of unpredictability which arise with neural firing
patterns will also contaminate these higher levels.  

It is at this point that problems of the philosophy of mind can merge
with those of the philosophy of physics.  If there are no facts except
observed facts, then what determines our observations is identical to
what our observations determine.  In this situation, we have to
combine ideas from two initially distinct fields of study.  One
consequence is that concepts of ``naturalness'' and of ``simplicity'' have
to be re-developed.  For example, it may be that what our awareness
determines and is determined by is not be something ``natural'' in
terms of elementary physics (perhaps involving particle positions or
the states of an orthonormal basis), but rather something ``natural''
and ``simple'' in terms of observation or awareness or information
while, at the same time, being ``natural'' and ``simple'' at some more
sophisticated level of quantum mechanical analysis.

\title{Patterns, Indicators, Manifestations.}
\endtitle

Natural correspondence between the level of neural information and
the quantum level requires some sophistication in analysis on both
levels.  For example, the idea of ``a pattern of neural firing'' is
certainly not sufficiently well-defined to correspond to a unique
pattern of quantum states.  A neuron is a macroscopic object which
firing takes a significant time to cross, and not all firings on a single
neuron are identical anyway.  Nevertheless, it is possible to find
spatially localized, quantum mechanically simple, subsystems of
neurons which can act as indicators of neural firing (Donald 1990,
1995, 1999), and it is possible to define ``patterns of neural firings'' in
terms of the spacetime arrangements of the neural functioning of
those indicators (Donald 1995).  And yet there does not seem any
natural way to make a unique identification of the indicators.  Not
only can they each vary in type but also they can each vary
continuously in, for example, position.  With my proposed definition
of a pattern, there are only finitely many patterns for a completely
specified set of indicators, and so it is possible to postulate a unique
correspondence between ``observer'' and pattern and to calculate
probabilities for observers in terms of that postulate.  Although each
pattern will still have continuously many ``manifestations'' as
quantum histories, corresponding to the possible variations in the
physical indicators, probabilities, patterns, and observers can be
defined over these continuous sets.

\title{The Coin Model.}
\endtitle

For a preliminary investigation, it is sufficient to continue with a naive
idea of a pattern of neural firings.  Suppose then that, as a first step,
we model such a pattern as a pattern of coin tosses in something like
the sort of game mentioned above.  Then the important information in
this model would be which coin (or neuron) was which and when it
turns (or fires).  This information can be defined in geometric terms,
by the position of the coins relative to others and by when each coin
turns relative to other tosses.  It is possible to abstract a finite pattern
from this information by considering only the spacetime
arrangements between the coin tosses, although, for any set of
arrangements, there are continuously many compatible positions
which actual coins could occupy.

\title{Biological Advantage.}
\endtitle

A physicist without much knowledge of neurophysiology might think
that most of the unpredictability in such a model would be caused by
uncertainty in external inputs to the brain.  An important purpose of
this paper is to point out that this idea is completely false.  In fact, the
precise order of neural firings is utterly unpredictable and most of
this unpredictability is due to the internal mechanisms of neural
functioning.  Unpredictability at a detailed level is entirely
comprehensible in biological terms.  Neural nets evolved in order to
co-ordinate rapid responses such as fight or flight.  It is vital that a
threatened animal flees from a predator, but it would give a great
advantage to the predator if the flight follows an exactly predictable
path.  The timing of turns and darts and lunges should always be
unpredictable.

\title{Sources of Unpredictability.}
\endtitle

There are many sources of fine-grained unpredictability in the brain. 
Indeed, the brain is a vast patchwork of metastable fluid systems with
the timing of each neural firing linked to that of many others.  As a
result, small uncertainties in the timing of one firing can rapidly
magnify to affect the entire macroscopic firing pattern.  Even in
recent work (Berry et al.\ 1997, de Ruyter van Steveninck et al.\
1997) in which the firings of certain sensory neurons in response to
given stimuli are demonstrated to be quite highly reproducible in
biological terms, such small initial uncertainties are present.  A major
source of uncertainty in the timing of firing comes from the
uncertainty of information passage between connected neurons.
Although there is a background of spontaneous firing, in general the
firing of one neuron is controlled (either enhanced or inhibited) by
the firing of the thousands of neurons which connect to it. 
Connections between neurons are made at synapses where electrical
signals are converted into chemical signals.  However, this conversion
is a stochastic process with a high failure rate.  Regehr and Stevens
(2001) review experimental evidence showing that, at least in certain
systems, the probability of an individual synapse making a given
conversion can average 0.3 with a mode of 0.15.  Once again,
randomness in individual synaptic connections is biologically
plausible.  It has long been believed that learning involves changes in
synaptic transmission.  As neural firing is to a large extent an
all-or-nothing effect, it is naturally controllable by alterations in the
probability of transmission.

Estimates vary, but there are probably at least $10^{14}$ synapses in
an average human brain.  Neurons fire at an average rate of order a
few times per second.  If every synaptic transmission is an uncertain
event with probability significantly distinct from 0 or 1, then there
will be at least $10^{14}$ such events per second in the brain.  Thus
uncertainty in ordinary everyday neural functioning may overwhelm,
by many orders of magnitude, many conventionally recognized
sources of observed ``quantum'' uncertainty and may, in fact, be the
major source of unpredictability in human affairs.  If we interpret
quantum states as states of knowledge, then any uncertainty about
what we do not know becomes a quantum uncertainty.  In classical
terms, whether a message is passed on by an individual synapse is
determined by the precise state of that synapse when the message
arrives; in particular, by the positions and states of the synaptic
vesicles which could disgorge their contents into the synaptic cleft.  In
quantum terms, the state of a synapse observed only by the effects of
its transmissions is a result of many individual unobserved  molecular
collisions.  It is a decoherent state with probabilities for many
different classical possibilities.  When a new transmission becomes
possible, the state of the synapse will be a mixture of that
transmission occuring and failing to occur.

Individual synaptic transmissions depend as much on small-scale
thermally driven molecular motions as on biologically significant
information.  And yet they can rapidly give rise to biologically
significant information, because they affect the timing of subsequent
neural firings, leading to greater differences at subsequent synapses,
and altering the complex pattern of combinations of firings and
feedbacks which are involved in every neural processing.  The brain
is warm and wet, unpredictable, unstable, and inhomogeneous.  There
is, in my opinion, no evidence that it functions in any way like a
coherent quantum computer, and nor can I see any plausible way in
which such functioning could have evolved.  However, the brain also
does not function like a deterministic classical computer which uses a
fixed mechanism to take given inputs into predictable outputs. 
Instead, the physical nature of the brain makes its outputs and its
detailed behaviour unpredictable.  At each moment, the short term
future of a brain has a very large number of different possible
macroscopic configurations, each with significant probability.

\title{How Large is the Space of Possibilities?}
\endtitle

$2^{(10^{14})}$ different patterns may seem high, but it is negligible
by \vadjust{\kern1pt}comparison with the minimum dimension
(around $10^{(10^{26})}$) of the space of thermally active
wavefunctions available to any quantum system with the physical
entropy of the brain.  These numbers are both measures of the
number of possibilities at the microscopic level.  Although the larger
should always be kept in mind in discussions which treat human
observations using concepts from simple quantum mechanical models,
even the smaller may seem far too large to be an appropriate indicator
of the lack of predictability at a mental level.  Indeed, although our
thoughts wander and we can surprise ourselves, they wander on
comparatively long timescales, while much of our mental content
seems to consist simply of received external data.  Nevertheless, as
with a random walk, it is not the rate of wandering but the dimension
of the space of possibilities which is ultimately most important in the
long term unpredictability.

In the short term, small variations in the space of patterns will
correspond to small variations in the precise pattern of excitation, and
particularly in the precise ordering, of the $10^{11}$ neural firings
which occur during a given second.  Although these variations may
not in themselves constitute functional distinctions -- for example,
Abeles (1991, chapter seven) suggests ways in which the same
messages might result from varieties of different initial local firing
patterns -- they do have the potential to seed functional distinctions. 
At the very least, this indicates that, in any theory based on
functional distinctions, all possible functional distinctions will be
genuine physical possibilities.  In other words, human possibilities --
whether, for example, someone might be about to say ``Um, well, it
seems to me'', or ``Well, it seems to me, um'' are like the possibilities
that a radioactive decay will cause a Geiger counter to make its next
click when its clock reads 2:03:17.434 or 2:03:17.412.

This seems to me to be a significant conclusion however one
interprets the idea of ``physical possibilities''.  Although our thoughts
naturally tend to be most concerned with features of our lives that we
can predict -- like what we should buy to eat for dinner tonight, or
when we should set off on a journey if we are to expect to get to a
given destination by a given time -- the unpredictable features
constantly push us around in the space of possibilities.  That space
has so many dimensions and the dynamics is so unconstraining that
after any deviation we should surely never expect to get back to
where we would have been.  The length of queue at the supermarket
and the mood in which we leave the shop depend on exactly how long
we take to make our choices and whether we happen to arrive just as
the till is being emptied.  The close miss around the corner as the idiot
boy-racer coming towards us overtakes, depends to the second on
when we left our house and when he left his.  Major historical
outcomes also turn on apparently minor choices and on precise
timings (Cowley 1999, Durschmied 1999) and so no doubt does the
genotype of our children.

The different possible orderings of neural firings demonstrate a
significant failure in the sometimes plausible analogy between neural
imaging and imaging on a computer screen.  Pixels on a screen are
refreshed in a determined order and once one frame is complete, the
next frame starts.  There is, by contrast, no rule which tells us where
the next neural event will be found.  The difference is important in
terms of the amount of information which might be ``known'' about
the underlying physical states. We cannot avoid this problem simply
by claiming that, regardless of the order of its formation, ultimately,
the same picture will always result, because this is false -- small
initial differences may lead to large final differences, and anyway,
there is no unambiguous way of choosing moments when ``frames''
might be said to be filled.  Thus, when it is suggested below that
instantaneous neural states might indeed be interpreted as pictures,
any individual change in a single pixel will determine a new
``instant''.  This seems almost inevitably to lead to the idea that the
timings of neural events need to be defined to sufficient precision that
changes in the time-orderings of each pair of spatially distinct
events can be distinguished.  But since this involves an ordering of,
say, $10^{11}$ events in a second, or at least an ordering of the
timelike separations among those events, this implies a temporal
precision which in biological terms is simply ridiculous.  Nevertheless,
although an appropriate minimum biological timescale for individual
local events might be $10^{-4}$s, a description, simple in physical
terms, of information changes across the brain does seem to require a
much shorter timescale.  The fact is that, like the minimum biological
neural lengthscale of $10^{-6}$m, a timescale of $10^{-4}$s is
``macroscopic'' in physical terms in this context.  To work on such
macroscopic scales directly would require the definition of some
method of averaging over the smaller scales and this would introduce
at least as much ambiguity as it would resolve.

\title{Counting Futures.}
\endtitle

If probabilities for our possible futures are to be defined, it has to be
possible to define those futures.  In Donald (1997), I argue in favour
of discrete probabilities.  Although continuous probabilities
can certainly be useful as models for circumstances in which
external observers can vary the scale of their observations over a
wide range, observers observe their own reality by direct experience
so that the scale of the observation is part of the experience and is not
variable.  Nevertheless, it is not easy to see how we should count the
number of possible ways in which, in a short period, we might have
knowledge of our own neural states.  Defining equivalence classes of
functional distinctions would require that we could say what it would
mean for two neural states to be functionally equivalent in such a way
that if neural state $A$ was equivalent to neural state $B$ and neural
state $B$ to neural state $C$ then neural state $A$ would be
equivalent to neural state $C$.  However, in general, if we work at a
high level, we are likely to see functional similarities between neural
states rather than functional equivalences.  Relations like ``similarity''
or ``closeness'' are not transitive and do not define equivalence
classes.  In my opinion, the simplest way to cope with this problem is
to invoke the fact that individual neural firings are effectively discrete
and to attempt to work with discrete indicators of those individual
firings.

\title{Finding the Observer.}
\endtitle

So, as a preliminary model of personal ``knowledge'', we can take the
changing faces of a set of $N$ coins, with $N$ at least $10^{11}$ 
(in Donald (1995), I suggest thousands of indicators per neuron,
making $N$ at least $10^{14}$).  As a first result, we have that, at any
moment, the Shannon entropy of the distribution of values for these
faces is quite large.  Suppose that we want to use such a model as a
foundation for the idea that quantum states are states of knowledge. 
As long as we assume that we can identify an observer and his
neurons and a set of neural status indicators and their current status,
this seems fairly straightforward.  For any given quantum system, the
quantum state to be assigned to that system should be the most likely
state given all that information, where by ``most likely'', we
presumably mean ``maximal entropy'' in some sense.  It seems to me,
however, that the assumption here is much too strong to initiate a
consistent foundation for quantum mechanics.  The difficulty of
finding a natural set of neural status indicators has already been
referred to, but, there is also the much harder problem of ``finding''
the observer.  A version of this problem will arise in any
indeterministic theory in which as soon as we take our eyes off an
observer he disappears into a soup of possibilities, but it can most
easily be expressed in the formalism of many-worlds quantum
theory.  In such a theory, if any implicit appeal to an external
observer is to be avoided, then the individual observer should pull
himself and his world out of the undifferentiated background state
which is the quantum state of the universe.  A state which is almost
entirely uncontaminated by observation is what we think of as the
``initial state of the universe''; in other words, the state at the big
bang.  (To make it entirely uncontaminated, we may need also to
allow variable big bang dates, and maybe even variable ``physical
constants'' (Donald 1999).)  As all states can be viewed as Heisenberg
states, this state can also be considered as a state for the present
moment, at which time it will be a superposition of all possible
current states for the universe given a big bang origin.  This is not a
state in which definite sets of neurons exist, waiting for information
to be painted into them.  Instead, sets of neurons only exist as
possibilities.  We must find the neurons in the process of finding the
information.

\title{Quantum Theory Needs An Interpretation.}
\endtitle

Interpreting coins as pixels, we can think of the instantaneous state of
the brain as a a picture; with each picture corresponding to a unique
choice from, for example, $2^{10^{11}}$ possibilities. It is then not
entirely implausible that, if we could give a precise definition of an
instantaneous three-dimensional structure as a ``neural snapshot'', we
could use those pictures to label our fundamental correlations, our
preferred basis, and our states of knowledge.  This might be possible. 
Is this the sort of idea which those who want to claim that quantum
states are states of knowledge are assuming?  Or are they actually
assuming that knowledge is somehow outside the realm of physical
law?  In either case, the claim that ``quantum theory needs no
interpretation'' seems to me to be obviously incorrect; in the first case
because the identification of fundamental structures is an
interpretative problem, and in the second because the identification
of the unphysical (``classical?'') realm is.

\title{The Problem of Temporal Progression.}
\endtitle

My own approach to this problem is somewhat different.  I attempt to
identify {\sl four-dimensional }structures by looking for developing
histories of neural snapshots.  This introduces technical problems in
the quantum mechanical analysis, because, in a conventional language,
it becomes necessary to allow for a history of apparent state
``collapses'' (i.e.~changes of Heisenberg states), rather than dealing
merely with a single apparent ``collapse'' out of the initial state of the
universe.  On the other hand, this does have the advantage of building
temporal progression into the structure.  The two major problems in
the interpretation of quantum theory can be seen as the ``preferred
basis problem'' -- or the problem of what quasi-classical entities can
exist at a moment; and the problem of temporal progression -- or of
how we go from one quasi-classical moment to the next.  Relativity
theory makes defining the notion of a ``moment'' a significant part of
these problems.  The idea that quantum states are merely clusters of
correlations fails to solve the problem of temporal progression over
more than a single step.  Of course we can use a single quantum state
to learn, for example, that if I have spin up then you will have spin
down; and even that, if we have spin up now then there is probability
${2 \over 3}$ that we will have spin down in five seconds.  However,
these are things which we learn only if we assume that we know
what ``we'' are and that we are outside the frame of the state, and
even then, we need to know how we should take account of our new
knowledge if we are to learn what might happen next.

\title{Localization and Number.}
\endtitle

Return to the ``neural snapshot'' picture.  What are the possible
futures of a given snapshot?  Are they the possible snapshots in the
same place?  ``Same place'' in which frame?  How can we accomodate
changes in the substructure of the localization due to relative motions
of biological components?  Do all possible future snapshots use the
same number of ``pixels'', or ``coins'', or ``neurons'', or  ``neural
indicators''?  How can we accomodate changes in number due to
growth or decay or turnover of biological components?  What number,
or level of detail, should we start with?  Issues of localization and
number, similar or related to these, seem to me to be at the heart of
the temporal progression problem.  Issues involving the localization of
a quasi-classical entity are tricky, because localization is alway tricky
in quantum theory, but the most difficult issues seem to be those
related to number.  Moreover, the unpredictability of the details of
neural processing means that alterations in number correspond to
significant changes in probabilities.  If most of the functioning of a
brain were predictable, then it would not make much difference to
short term future probabilitities if there was a change of the level of
coarse-graining under which neural structure was defined.  As it is,
however,  it seems plausible that, in the language of a many-worlds
interpretation of quantum theory, most of the ``worlds'' we
experience differ by differences due to events within our heads
rather than due to external events, so that if we want to count our
worlds -- which should be the first step in defining probability --
then we have to be able to identity neural ``events''.

\title{The Trimming Problem.}
\endtitle

My analysis of this situation has eventually led me to define a
finite Markov process on a space of patterns of histories of neural
events.  In terms of such a process, the problems which need to be
solved in order to explain the experience of an indeterministic physics
are to identify a state space and to define transition probabilities on
that space.  The prevalence of neural unpredictability means that
transition probabilities for such a process can be strongly dependent
on the precise definitions used.  In particular, a specific number
problem arises which, in Donald (1995), I refer to as the ``trimming
problem''.  In the coin model, with a space of $10^{11}$ coins, the
problem would be that if a few of the coins were omitted from the
state space, then probabilities would increase but hardly any
significant information would be lost.  This is troubling, because even
in a classical picture of a brain, alterations in neural substructures
happen all the time while, as has already been discussed, in a fully
quantum mechanical picture, the neurons themselves only exist as
possibilities.  In Donald (1999), by detailed analysis of the temporal
structure of an observer, I propose a solution to this problem which
allows ``coins'' or ``neural indicators'' or ``switches'' to be added
retrospectively.  This allows the structure of an observer to develop
towards a natural balance between increase of number and decrease
of probability.

\title{An Abstract Approach.}
\endtitle

The difficulties of finding neurons in the initial state of the universe
leads me to an entirely abstract approach to the definition of a
``pattern''.  I do not directly define a ``pattern of neural firing'', but
rather a pattern of a certain type of information coded as spacetime
arrangements of ``yes-no'' events (``switchings'').  The yes-no events
correspond to the heads and tails of the coins or to the firings and
not-firings of the neurons.  The spacetime arrangements define
whether pairs of different events are spacelike separated or time
ordered.  The total amount of information in a pattern is finite, but
because the spacetime relations are defined for every pair of events,
a considerable amount of biologically irrelevant information is given,
and the space of patterns available to a single observer is
correspondingly very large.  The resulting plurality might be
thought of as the price of a simple abstract structure, although it turns
out that it does have advantages (Donald 1999, section 8).  The
primary purpose of this paper, however, is not to repeat my earlier
work; the significant points here are simply an indication of the type
of problems which may arise when one tries to express human
knowledge as a pattern of physical events, and the comment that an
attempt to solve the preferred basis problem locally may well lead to
a ``solution'' in which distinct possibilities are not actually defined by
orthogonal wavefunctions.  Thus, with my definition, pairs of different
patterns will often correspond to overlapping rather than orthogonal
states.  This does complicate the analysis of probabilities, in particular
in that it introduces a normalization requirement, but while
orthogonality may seem a reasonable criterion in the context of
simple models in low dimensional Hilbert spaces, it is of little
relevance for thermal states of localized macroscopic objects, for
which wavefunction descriptions are in general inappropriate.  The
very name ``preferred basis problem'' begs the question of whether a
basis is the right way to classify distinct possibilities at the level of
the human observer.  In my opinion, it is not.

\title{Public and Private Physics.}
\endtitle

It is tempting to think that all this is just very esoteric and to claim
that events external to an observer are not affected by internal
events and that therefore physics can be restricted to the calculation
of probabilities for those external events.  The preliminary model here
would be that we have two systems of coins some of which constitute
an observer and some of which are external.  Our modelled restriction
of physics would then be permissible as long as we could show that
the external coins are independent of the internal coins.  This is
certainly plausible and, under this sort of assumption, we can
construct a common, public, intersubjective physics.  But the question
we ultimately have to face as individual observers is why, as separate
individuals, this public physics is also the physics that we observe
privately.  The complexities of neural functioning outlined in this
paper make this not an entirely simple question to answer.  

Observer-effects might arise in many ways in a system as complex
as a human brain, and we would, perhaps, like to be able to
argue, for example, that no such effects arise due merely to increased
neural activity.  However, Donald (1999) provides an example of just
how difficult it can be in a fully-defined theory to argue for the
general absence of observer-effects.  In that paper, I split the
problem of relating public physics to private physics into two theses. 
The first thesis states that a typical modern human observer should
be aware of a world in which quantum theory is accepted and in
which its detailed theoretical predictions are confirmed.  The
argument for this thesis considers the observation of summaries of
many experimental results.  Classical and quantum versions of the
laws of large numbers can be invoked to tell us that the theoretical
probability that such summaries are consistent with quantum theory
should be close to one.  That events of theoretical probability close to
one should typically be observed is a comparatively simple test for a
model of observation.  The second thesis, however, is the claim that,
under appropriate circumstances, there should be a fairly direct
agreement between public and private probabilities for observing
single individual quantum events; for example, individual clicks of a
Geiger counter.

Intuitively, there would appear to be no problem in calculating these
probabilities:  Consider the possible events in a given time interval. 
Divide those events into a class in which a click is heard and a class in
which a click is not heard.  Weigh the counting of each event by its
individual quantum probability.  Use the weighted sum over the two
classes to give the probabilities of hearing and of not hearing a click
in the given interval.  

This intuitive picture, however, is, once again, a picture from the point
of view of an implicit external observer.  The nature of classes of
events and of fixed time intervals are already decided when we
choose to imagine a world which is like the world we see but
indeterministic.  But what we should be trying to imagine is a world
in which the observers are self-defining.  The division into classes of
events should be made internally rather than externally.  Everything
which makes a click-hearing observer a click-hearing observer
should be in the structure of the relevant quantum states and their
temporal development.  Now we really do face problems in saying
what an event is which are analogous to the preferred basis problem
or the consistent-histories set selection problem.  And we do have to
solve the trimming problem.  The events which ultimately lead to an
observation are neural events rather than external events.  Neural
observations are made by the parallel processing of many small
pieces of ambiguous information over time intervals which allow
many successive firings at some individual neurons and therefore a
combinatorial explosion of possible patterns of synaptic events over
the entire brain.  This suggests that observer effects might even arise
due simply to the time taken for a particular outcome to be realized.

These are problems which should not be ignored.  If they are
addressed, then they can be used to constrain interpretative
hypotheses.  For example, there was a significant change between the
proposals of Donald (1995) and of Donald (1999).  This was required
precisely in order to avoid an implausible observer-effect in the
situation of listening to a Geiger counter.  The difficulty stems from
the fact that a click naturally excites more neural activity than
silence.  The proposal of Donald (1995) failed because it ignored
``non-events''; where a coin face changing would model an event,
while a coin face which stayed unchanged would model a non-event. 
The normalization of the probabilities of overlapping states could
then lead to considerable under-weighting of neural inactivity.  The
solution I put forward in Donald (1999) effectively allows
non-activity to be observed in circumstances in which activity can be
observed.  The details here may only be important in the specific
context of my proposals, but there should be no doubt that, in any
kind of detailed analysis, problems will arise  from the underlying
facts that neural states are thermal and that they have much more
complexity over short time scales than simple models might suggest. 
A cat being looked at by a person is not something  being seen simply
as ``dead'' or ``alive''; rather it is one enormously complex quantum
system interacting through several different channels with another
even more enormously complex quantum system.

\title{Avoiding the Problems.}
\endtitle

Technical problems arise as soon as one accepts the goal of describing
personal observations by a well-defined stochastic process.  In order
to avoid those problems, one might argue that such a description is
unnecessary.  For example, one might claim that it is not necessary for
probabilities of single events to be precisely defined; or that an
observer's self-knowledge is never perfect; or that memory is part of
the instantaneous present structure of an observer so that there need
be no well-defined connection between the past of an observer and
the present; or that observers simply can exist in innumerably many
ways and that our experience just happens to be of one of those ways;
or finally, one might claim that some alternative physical theory
makes any problems about the nature of observers irrelevant.

\title{Are Probabilities Precise?}
\endtitle

The claim that it is not necessary for probabilities of single events to
be precisely defined is surely correct for several of the ways in which
probabilities have been understood (Gillies 2000). For example,
prominent among these ways is the suggestion that probabilities are
relative frequencies; applying not to single events, but to infinite
sequences of events.  Alternatively, if probabilities for individual
events are characterized by betting quotients or by degrees of belief,
it would also seem unnecessary that they should be exact; at least,
as long as the idea of a belief is itself not taken to be foundational for
physics.  Nevertheless, the fact that there are situations in which these
kinds of imprecise probabilities are appropriate and useful, does not
rule out a propensity theory for an indeterminstic physics in which
there are some fundamental probabilities which are physical facts
defined as part of the physical laws.  A propensity theory is also not
ruled out by the fact that, because probabilities can only be measured
by looking at long sequences of events, their measurement can
never be exact.  Indeed, it seems to me that the probabilities in any
physical theory with fundamental indeterministic events have to be
lawfully-defined propensities.  One might, for example, interpret such
propensities by supposing that a random string of digits is part of the
initial conditions of the universe, and that it is part of the laws of
physics that digits from the string should be brought into play at
appropriate moments.  If the probability of a future event is a
fundamental physical fact of this kind, then the precision which is
required to make such a theory complete is precision in the lawful
way in which randomness is brought into play.  The interpretation I
propose does have this precision.

\title{Epistemology or Ontology?}
\endtitle

It is, of course, correct to say that an observer's self-knowledge can
never be perfect.  My counter-claim is that ``observers'' are
fundamental parts of our reality and exist as definite entities.  As I
have no idea what vague existence can mean, unless it is existence
through the eyes of an external observer, I believe that this amounts
to no more than taking the position on the mind-body problem that
``minds'' exist.  Indeed, to the extent that it makes observers
fundamental, my developed theory is close to a form of idealism, with
physical laws and initial conditions merely providing probabilities for
mental histories.  Well-defined probabilities for observations and
definite existence of observers are the mutually sustaining demands
by which I have found myself led, much to my surprise, towards this
idealism.

\title{Escapism.}
\endtitle

The claims that there need be no direct connection between past and
present and that we might exist just in one of innumerably many
ways seem to me both to be absurd.  They are absurd in the same
sense that the conceit that one might be a brain in a vat is absurd. 
Philosophical scepticism is important.  We should always be aware of
how little of our knowledge is certain and be prepared to doubt
everything.  But beyond a certain point, the idea that anything is
possible becomes escapism.  Above, I describe the problem of
temporal progression as one of the two major problems of the
interpretation of quantum theory.  I think that it is escapism to claim
that this problem can be ``solved'' by ignoring it and supposing that
we have memories but not pasts.  Carried to its logical extreme,
this produces the timeless theory of Barbour (1999).  In such a theory,
predictive probabilities are illusory; our probability calculations are
merely present memory traces of probability calculations; our hopes
and fears are empty.  Any plausibility this idea might have as a theory
of mind, would seem to depend on some sort of materialist
functionalism in which a mind is taken to inhere in the way that an
instantaneously associated brain would function were there time in
which it could.  In Donald (1997), I have criticized the attempt to base
functionalism on the idea of disposition to function rather than actual
historical functioning.  Where there is never any actual historical
functioning, however, the very idea of a disposition to function seems
to me to be ludricous. 

As to the suggestion that our experience just happens to be one of
innumerably many possibilities, this seems to me even more dubious
than the suggestion that we only exist in the present.  Instead of
Barbour's explicit statement that we have no past and no future, we
would be encouraged to believe in past and future, but rather than
being required to construct them, we would be supposed to allow in
some unexplained way for all possible frameworks.  Thus we would be
supposed to contemplate ourselves with futures and pasts defined
some arbitrary path through the states, sometimes of 6,  and
sometimes of 6,000, and sometimes of 4,721,637, of our neurons. 
Once again, I suspect that this idea ultimately depends on materialist
functionalism, as how else are we supposed to make sense of any
particular collection of neurons except by trying to work out how that
collection either might behave or has behaved?  Once again, however,
any form of functionalism seems questionable without a clear choice of
dynamical framework.

\title{Hume and Parfit.}
\endtitle

Following Hume (1739), Parfit (1984) describes persons as being like
``nations''.  A preliminary version of this paper was presented
in Belfast; a reminder of the extent to which a nation is a
purely human construct, with arbitrary boundaries drawn and
redrawn by a succession of historical accidents (Jackson 1997).  I
do not disagree either with Parfit or with Hume about the ephemeral
nature of experienced personal identity, but instead suggest that
without a basis for that experience through the existence of a
definable observer, a person would be as indefinite as, without an
imposed legal framework, Belfast would be nationless.  Hume suggests
that personal identity amounts to nothing more than ``a succession of
parts, connected together by resemblance, contiguity, or causation'',
while Parfit claims that ``a person's existence just consists in the
existence of a brain and body, and the occurence of a series of
interrelated physical and mental events''.  My problem with these
views is that, in the context of a global quantum theory, and without
assuming an external observer, I find it difficult to give meaning to
the terms  ``succession'', ``parts'', ``connected together'', ``resemblance'',
``contiguity'', ``causation'', ``brain'', ``body'', ``series'', ``interrelated'',
and ``events''; let alone, ``physical'' and ``mental''.  Hume and Parfit
could point to a ``body'' as the approximate or temporary possessor of
an identity.  Neurophysiological detail makes the external behaviour
of a body an insufficient guide to its internal experience, while
quantum mechanics calls any act of pointing into question.  A
definition for an observer might thus be seen as the least which is
required for a precise reformulation of those problematic terms. 

Many of Parfit's arguments about personal identity  are based on
science-fiction thought experiments. It is possible to analyse each of
these experiments in the light of the definition proposed in Donald
(1999) and, at least when sufficient imaginary details are added, it is
possible to draw definite conclusions.  For example, I would always
recommend travelling to Mars by spacecraft rather than by atomic
disassociation on Earth and reconstruction on Mars.  From my point of
view, Parfit's examples tend to turn on ways in which the
abandonment of naive functionalism can open a gap between observed
and experienced behaviour.  Any specific definition of personal
identity in terms of physical structure implies the possible existence
of things which are actually not conscious but which appear to behave
as if they were.  It is almost inevitable, therefore, that problematical
thought experiments can be described.  According to my definition,
consciousness depends on a physically continuous past history.  If my
body were to be reconstructed on Mars, I would argue that the actual
consciousness of that body would be consciousness of a history
beginning at the moment of reconstruction.  Of course, being
dependent in behaviour entirely on its physical composition, it would
only be able to behave, as far as external observers are concerned, in
ways in which my current, physically identical, body would behave. 
Thus it would say that it did not necessarily believe what it would
tend to say if it tried to describe what it thought it was feeling!  But
the question we should be asking here is not, ``what is that person
saying about his experiences?'', but rather, ``what would it be like to
be nothing but an experiencing of that middle-aged body over just a
few minutes, hours, or days?''.  From the point of view of an external
observer, this might seem to make ordinary consciousness look like a
miraculous pre-established harmony between physical behaviour and
awareness, because there is no necessary link between the behaviour
seen by the external observer and awareness.  However, the miracle
lies rather in the existence of laws which can make long, rich,
meaningful, coherent histories plausible.  With suitable laws in place,
the reason that there would tend to be harmony between physical
behaviour and awareness, would be that awareness would be far
more likely to arise in the context of appropriate behaviour than
otherwise.

\title{The Now.}
\endtitle

Saunders, Wallace, and Butterfield all emphasize the analogy between
splittings of a universal state into worlds and splittings of a spacetime
into instants.  Butterfield (2001), for example, writes ``just as
someone who accepts the tenseless conception of time can readily
accept instants i.e.~spacelike slices of spacetime, as (i) useful or even
indispensable for describing phenomena, and yet (ii) not any
substantive ontological commitment additional to the commitment to
spacetime; so also an Everettian can readily accept worlds as (i) useful
or even indispensable, and yet (ii) not a substantive commitment
additional to the commitment to actuality's being described by the
universal state.''  While ``instants'', considered purely as an arbitrary
mathematical slicing, may not imply a substantive additional
ontological commitment, I would rather make the analogy between
 the Everett ``world'' that one is seeing and ``now'' -- the time that one
is seeing.  In these terms, the analogy underlines the ontological
commitment to a ``one'' or an observer.  Ultimately, defining the
present instant for an observer is equivalent to defining the present of
the observer or to defining the observer's present world.

\title{Observer Relative Terms.}
\endtitle 

The fundamental problem which the idea of psycho-physical
parallelism raises for a physicist is whether anything needs to be
added to the mathematical formalism of a physical theory in order to
understand the apparent existence of observers.  The standard
justification of the proposal that nothing needs to be added seems to
depend on the argument that there just are some natural physical
structures which behave in particularly complex ways.  As I see it,
however, ``natural'', ``structure'', ``behave'', and even ``complex'', are
all in fact observer-relative terms (cf.~Searle 1992, 9.V), and so this
leads to the problem of whether such structures can be, in some way,
reflexive, self-relative, or self-defining and if so, whether they are
sufficiently well-characterized by themselves to specify their
individual pasts and their possible futures.  This seems so implausible
to me that I am prepared to propose as an alternative that there have
to be fundamental physical laws defining the existence of observers
(Donald 1999).  The frequent invocation of assumptions about
decoherence might be thought of as attempts to postulate such laws
without having to make them explicit.  In my opinion however, these
postulates are insufficient.  Decoherence merely helps to provide the
circumstances out of which observers can emerge.

\title{Alternative Physical Theories.}
\endtitle

At least implicitly, the idea of the observer is central both to the idea
that quantum states are states of knowledge and to the idea that
quantum states are clusters of correlations.  Given that there are
difficulties in characterizing the physical structure of observers,  it
might seem appropriate to pursue some of the alternative
approaches to the interpretation of quantum theory in which
observers are not fundamental.  At present, however, I am unaware of
any such approach being fully compatible with experiment, with
special relativity, and with quantum field theory.  But even if we did
have such an interpretation, the problems raised in this paper would
still be important.

\title{Ignorance Probabilities.}
\endtitle

In a many-minds theory, the fundamental indeterministic events are
individual observations.  In most other theories, the probabilities of
observations for an individual are thought of as ``ignorance
probabilities''.  The conventional use of such probabilities involves
supposing that an observer's direct information can be combined with
theoretical knowledge in order to provide rational odds for future
observations.  For example, a Maxwell distribution may be predicted
for molecular velocity observations as soon as the temperature of a
gas at equilibrium has been observed.  Probabilities of this type are
not fundamental and need not be precisely defined for individual
events.  According to a classical theory, for example, the gas molecules
simply have some definite velocities, which result from their past
collisions and which cause thermometers to display the appropriate
temperature.  But although each individual measurement is simply an
isolated fact, sufficiently long sequences of velocity measurements
will be likely to fit a Maxwell distribution, or some version of the
distribution corrected for a more sophisticated theory, to any
required degree of accuracy.  We can understand this theoretical
distribution as being an appropriate description of any of the vast
majority of ways in which a certain total energy can be shared by
collisions between molecules, but we can hardly expect any actual gas
sample to achieve perfection in randomization.  It is none the worse
for this type of ignorance probability that it is just an expression of
the somewhat circular idea that we should expect to see a typical
situation.  Theoretical progress can still come from our ability to
improve our description of ``typical situations''.  Conceptually, we can
say that the situation which we happen to be seeing is just a
reflection of the initial state of the universe, and the fact that that
situation is typical merely tells us that no special explanation of this
aspect of the initial state need be given.

\title{Including the Observer.}
\endtitle

As always, however, problems arise when we try to include the
observer in the situation to be described.  Now the fact that what we
see depends on what we are enters into the equation.  The focus of
this paper has been on the probabilistic prediction of the future
development of the information which defines consciousness given the
past history of that information.  The neural unpredictabilities and
instabilities discussed above make even short-term probabilistic
predictions of this kind highly dependent on precisely what form the
information takes, regardless of whether the probabilities in the
predictions are ignorance probabilities or propensities.  Indeed, if the
information includes precise positions for all the atoms in the brain,
then short-term prediction of future firing patterns will clearly be
much more reliable than if the information merely gives the past
history of firing patterns.  However, unlike in a many-minds theory
or in a theory of quantum states as states of knowledge or of quantum
theory as a theory of correlations, it can be argued in the context of a
theory involving ignorance probabilities that being able to step from
present information to future information is not fundamental.  What
we might see and how likely we are to see it, is not caused, but
merely filtered, by what we are.  Whatever will be will be.  It is just a
fact that we do not and cannot know what the future will bring and it
is just another fact that we do not and cannot know exactly what we
ourselves are. Nevertheless, this does not imply that there is not
something which we exactly are.

\title{External and Internal Observers.}
\endtitle

From the point of view of an external observer looking at a functioning
human brain, the physical nature of human mental processing makes
it seem difficult to identify anything as the essence of the existence of
that brain for itself.  However we are also, and indeed primarily,
internal observers looking out.  From that point of view, it seems
necessary to accept that we are something.  And if we are something,
then, given the absurdity of vague existence, there is something
which we exactly are.  The contrast between internal and external
points of views is reflected in a contrast between physical theories. 
On the one hand, there are theories, like many-minds theories, in
which the concept of information is fundamental; and on the other
hand, there are theories in which external physical reality is
fundamental.  In the first type of theory, the appearance of external
physical reality is derived from the rules which determine the kind of
information which exists; but in the second type of theory, the
existence of information for itself seems rather unnatural.

\title{Advantages and Disadvantages.}
\endtitle

In an information-based theory, fundamental information can be
taken to be local because an individual's information exists at that
individual; a clear and simple recent analysis of this point is given by
Fuchs (2002, section 3).  This makes it possible to avoid the locality
problems which plague other interpretations.  Fundamental
information can also be taken to be information possessed by
individual observers.  This puts mind at the heart of our physical
theories and reflects our individual realities.  Nevertheless, the
advantage is not all in one direction.  As we have seen, in an
information-based theory in which local private information is
fundamental, there is the possibility of ``observer effects'' which could
make the likely private observations of individuals differ from
textbook predictions of public observations.  Except in cases with
significant changes in the actual structure of the observer as a result
of particular observed events, this problem does not arise with a
theory involving ignorance probabilities, because, with such a theory,
private observations are a consequence of public events.  Thus, for
example, with an ignorance theory, we hear a sequence of clicks
which really happened, rather than, with a theory in which there are
no facts except observed facts, becoming a hearing of a sequence of
clicks.  If, at the end of the sequence, we intend to publish a
statistical summary of the clicks which we have heard, then that
summary will have the same status for us as for anyone else,
according to an ignorance theory.  Under a many-minds theory,
however, it will be a private event for us, but a public event for any
other observer.

\title{Classical Deterministic Theories.}
\endtitle

Even in terms of nineteenth century physics, the idea that an observer
is a physical subsystem which behaves as an observer would not
be sufficient to delineate a unique class of human observers because
the question of the scale at which a mind experiences a body would
remain open.  In such a theory, an attempt to understand
psycho-physical parallelism would start with the idea that a given
mind is an awareness of the workings of a given body.  ``Given'' is a
give-away for an implicit external observer, but even if it is allowed
that, in a classical deterministic theory, ``bodies'' are natural physical
structures, it is not at all clear that their ``workings'' are equally
natural.  It might be, for example, that an observer with awareness
defined at the molecular level would be different from an observer
with awareness defined at the neural firing level.  While one could
say that it would be unreasonable to suppose that an observer would
have awareness defined, in an arbitrary way, sometimes at the
molecular level and sometimes at the neural firing level, this sort of
remark is perhaps as far as one can go in a classical deterministic
theory, towards constructing a reasonable and consistent picture of a
human observer.

\title{The Bohm Interpretation.}
\endtitle

The Bohm interpretation provides a deterministic model of
non-relativistic quantum mechanics.  There are significant problems
in understanding psycho-physical parallelism in this theory, in as far
as it is not clear what role the hidden variables (the particle
positions) should play (Albert 1992, Page 1995, Brown 1996).  Particle
trajectories are determined by the positions of other particles and by
the guiding wave-function, and the physical constitution of an
observer could depend on facts about both.  Indeed, it could even be
that the particle positions are entirely irrelevant to psycho-physical
parallelism and that, despite the existence of those definite positions,
we are aware of a stochastic process of patterned structures in the
global guiding wave (which is Everett's universal wave-function). 
The spirit of the Bohm interpretation, however, would seem to
suggest that we are aware, at any moment, either of the positions of
some family of particles associated with our brain (Bell 1981), or that
we have just enough information about some such family to be able to
assign them an effective local wavefunction (Albert 1992).  And yet it
is unclear which family of particles is involved, nor how that family is
updated as the brain changes.  Moreover, the information is of a type
which seems quite remote from biologically significant information. 
In my opinion, this makes the Bohm interpretation an example of a
theory in which the existence of consciousness, although possible, is
hardly ``natural''.

\title{Stochastic Theories.}
\endtitle

Stochastic or indeterministic theories  -- such as the spontaneous
collapse theories reviewed by Stamatescu (1996) and by Pearle
(1999) -- can be interpreted in two distinct ways.  In the first way,
the idea is that the indeterminism is a matter of ignorance so that the
situation is like that of classical statistical mechanics, or even Bohmian
mechanics, with some ultimate deterministic theory underlying and
causing the apparently indeterministic events.  In this case, the
theories present similar problems to the Bohm interpretation, only
made worse in that the underlying deterministic theory is unknown. 
The alternative interpretation of stochastic theories is to suppose that
the probabilities are objective.  In this case, if there is something
which we exactly are, then this something will also change
indeterministically and it might indeed be conceptually accurate to
see ourselves as something like some subset of a set of  coins being
tossed.  There being something which we exactly are would now
require rules as to which subset we are and as to how that subset
might alter over time with differing possible outcomes.  In both cases,
as with the Bohm interpretation, it may seem unnatural that
awareness should be defined in terms of structures at the level of the
underlying physical theory, in as far as those structures seem so
remote from the level of biologically significant information. 

\title{Quasi-classical Theories.}
\endtitle

Collapse theories and the Bohm interpretation are both motivated by
the goal of constructing a quasi-classical physics in which, in
particular, macroscopic objects can be said to have quite well-defined
positions at all times.  As well as problems with locality, theories
which aim at a quasi-classical physics also tend to have more general
problems with quantum field theory, according to which the same
physical systems can have both quasi-classical particle-like states
and quasi-classical field-like states.  This makes it difficult to see
how one can make a universal choice of fundamental quasi-classical
variable.  Moreover, in order to explain our observations, we do not
need to construct a quasi-classical physics; what we need to do is to
explain our observations.  That remains a problem even given a
quasi-classical physics for the world external to an observer.

\title{Conclusion.}
\endtitle

The ultimate purpose of theoretical physics is to provide a consistent
and plausible theoretical framework for our individual observations,
including what we learn from others.  However, the detailed nature of
our individual observations depend on precisely what parts of the
workings of our brains form the physical aspect of our
consciousnesses.  In approaches to the interpretation of quantum
mechanics in which the ideas of information or correlation or mind
are fundamental, there are no events except observed events, and
those events and their possible futures are defined by the solution to
this problem of psycho-physical parallelism.  However the
unpredictability of the normal functioning of the human brain means
that different solutions at different scales will involve different sets
of events with different future probabilities which cannot be assumed
to be better than approximately consistent.  Thus without some
solution to the problem of psycho-physical parallelism such
interpretative approaches are incomplete.  Nevertheless, solving the
problem is difficult.  Some of the ways in which difficulties can arise
have been sketched.  Finally, it has been suggested that, like the
classical physics they attempt to emulate, some alternative
approaches to the interpretation of quantum mechanics provide
frameworks in which human consciousness seems a peculiar and
extraneous appendage to reality.

\proclaim{Acknowledgements.}{}
\endproclaim

I am grateful to both Chris Fuchs and David Wallace for stimulating
walks and challenging conversations.

\proclaim{References.}{}
\endproclaim

\frenchspacing
\parindent=0pt

\everypar={\hangindent=0.75cm \hangafter=1} 

Abeles, M. (1991) {\sl Corticonics -- Neural Circuits of the Cerebral
Cortex.}  (Cambridge)

Albert, D.Z. (1992) {\sl Quantum Mechanics and Experience.}
(Harvard)

Barbour, J. (1999) {\sl The End of Time.} (Weidenfeld and Nicolson)

Barrett, J.A. (1999) {\sl The Quantum Mechanics of Minds and
Worlds.} (Oxford)

Bell, G.H., Emslie-Smith, D., and Paterson, C.R. (1980)  {\sl Textbook of
Physiology.} 10th edition (Churchill Livingstone) 

Bell, J.S. (1981)  ``Quantum mechanics for cosmologists.''  pp
611--637 of {\sl Quantum Gravity 2,} ed. C. Isham et al.~(Oxford). 
Reprinted, pp 117--138 of Bell (1987).

Bell, J.S. (1987)  {\sl Speakable and Unspeakable in Quantum
Mechanics.} (Cambridge)

Berry, M.J., Warland, D.K., and Meister, M. (1997), ``The structure and
precision of retinal spike trains.''  {\sl Proc. Natl. Acad. Sci. USA \bf
94}, 5411--5416.

Brown, H.R. (1996), ``Mindful of quantum possibilities.''
{\sl  Brit. J. Phil. Sci. \bf 47},  189--200.

Butterfield, J.N. (2001) ``Some worlds of quantum theory.'' 
{\sl quant-ph/0105052}.  A version of this paper is to appear in a
CTNS/Vatican Observatory volume on Quantum Theory and Divine
Action, (ed. R. Russell et al.).

Caves, C.M., Fuchs, C.A., and Schack, R. (2001)  ``Quantum probabilities
as Bayesian probabilities.'' {\sl quant-ph/0106133}

Chalmers, D.J. (1996)  {\sl The Conscious Mind.} (Oxford)

Cowley, R. (1999)  {\sl What If?} (Putnam)

Dennett, D.C. (1991) {\sl Consciousness Explained.} (Little Brown)
\smallskip

Donald, M.J. (1990) ``Quantum theory and the brain.''  {\sl Proc.
R. Soc. Lond. \bf A 427},  43--93.

Donald, M.J. (1992) ``A priori probability and localized observers.'' {\sl 
Foundations of Physics, \bf 22}, 1111--1172.

Donald, M.J. (1995) ``A mathematical characterization of the physical
structure of observers.'' {\sl  Foundations of Physics \bf 25},
529--571.

Donald, M.J. (1997)  ``On many-minds interpretations of quantum
theory.'' \newline {\sl quant-ph/9703008}

Donald, M.J. (1998)  ``Discontinuity and continuity of definite
properties in the modal interpretation.''  In {\sl The Modal
Interpretation of Quantum Mechanics} (ed. D.
Diecks and P.E. Vermaas),  pp 213--222.  (Kluwer)

Donald, M.J. (1999)  ``Progress in a many-minds interpretation of
quantum theory.'' {\sl quant-ph/9904001}

{\hfill My papers are also available from \quad {\catcode`\~=12
\catcode`\q=9 http://www.poco.phy.cam.ac.uk/q~mjd1014} \hfill}
\smallskip

Durschmied, E. (1999) {\sl The Hinge Factor.} (Hodder and Stoughton)

Fuchs, C.A. (2001a)  ``Notes on a Paulian idea: foundational,
historical, anecdotal and forward-looking thoughts on the
quantum.''  {\sl quant-ph/0105039}.

Fuchs, C.A. (2001b)  ``Quantum foundations in the light of quantum
information.''  To appear in {\sl Proceedings of the NATO Advanced
Research Workshop on Decoherence and its Implications in Quantum
Computation and Information Transfer} (ed. A. Gonis).
{\sl quant-ph/0106166}.

Fuchs, C.A. (2002)  ``Quantum mechanics as quantum information (and
only a little more)''  A revision of Fuchs (2001b).
{\sl quant-ph/0205039}.

Fuchs, C.A. and Peres, A. (2000)  ``Quantum theory needs no
`interpretation'.''  {\sl  Physics Today}, Article: March, 70--71.
Correspondence and Replies: September, 11, 12, 14, 90.

Gillies, D. (2000)  {\sl Philosophical Theories of Probability.}
(Routledge)

Griffiths, R.B. (1998) ``Choice of consistent family, and quantum
incompatibility.'' {\sl Phys. Rev. \bf A 57}, 1604.  {\sl
quant-ph/9708028}.

Hume, D. (1739) {\sl A Treatise of Human Nature, Book I: Part IV:
Section VI.} (Noon)

Jackson, A. (1997)  ``British Ireland.''  In {\sl Virtual History} (ed.
N. Ferguson),  pp 175--227.  (Picador)

Mermin, N.D. (1998) ``What is quantum mechanics trying to tell us?'' 
{\sl Amer. J. Phys. \bf 66}, 753--767. {\sl quant-ph/9801057}.

Page, D.N. (1995)  ``Attaching theories of consciousness to Bohmian
quantum mechanics.''  To appear in {\sl Bohmian Quantum Mechanics
and Quantum Theory:  An Appraisal}, (ed. J.T. Cushing, A. Fine, and
S. Goldstein) (Kluwer, 1996).  {\sl quant-ph/9507006}.

Parfit, D. (1984) {\sl Reasons and Persons.} (Oxford)

Pearle, P. (1999)  ``Collapse models.''  To appear in {\sl Open Systems
and Measurement in Relativistic Quantum Theory}, (ed. F. Petruccione
and H.P. Breuer) (Springer, 1999).  {\sl quant-ph/9901077}.

Peierls, R. (1991) ``In defence of `measurement'.'' 
{\sl Physics World}, January 19--20.

Regehr, W.G. and Stevens, C.F. (2001) ``Physiology of synaptic
transmission and short-term plasticity.''  Chapter three of W.M. Cowan,
T.C. S\"udhof, and C.F. Stevens, {\sl Synapses.} (John Hopkins)

Rovelli, C. (1996), ``Relational quantum mechanics.'' {\sl Int. J. Theor.
Phys. \bf 35}, 1637--1678. {\sl quant-ph/9609002}

de Ruyter van Steveninck, R.R., Lewen, G.D., Strong, S.P., Koberle, R.,
and Bialek, W. (1997) ``Reproducibility and variability in
neural spike trains.''  {\sl Science \bf 275}, 1805--1808.

Saunders, S. (1995) ``Time, quantum mechanics, and decoherence.''
{\sl  Synthese, \bf 102},  235--266.

Saunders, S. (1998) ``Time, quantum mechanics, and probability.''
{\sl  Synthese, \bf 114},  373--404.

Searle, J.R. (1992)  {\sl The Rediscovery of The Mind.} (Bradford)

Stamatescu, I.-O. (1996)  ``Stochastic collapse models.'' Chapter eight
of D. Giulini, E. Joos, C. Kiefer, J. Kupsch, I.-O. Stamatescu, and H.D.
Zeh, {\sl Decoherence and the Appearance of a Classical World
in Quantum Theory.}  (Springer)

Wallace, D. (2001a)  ``Worlds in the Everett interpretation.'' {\sl
quant-ph/0103092}

Wallace, D. (2001b)  ``Everett and structure.'' {\sl
quant-ph/0107144}

Wheeler, J.A. (1957)  ``Assessment of Everett's `relative state'
formulation of quantum theory.''  {\sl Rev. Mod. Phys. \bf 29},
463--465. 

Wigner, E.P. (1961)  ``Remarks on the mind-body question.''  In {\sl
The Scientist  Speculates} (ed. I.J. Good), pp 284--302. (Heinemann)

Wolfe, H.C. (1936) ``Quantum mechanics and physical reality.'' 
{\sl Phys. Rev. \bf 49}, 274.

\end